\documentclass{ws-mpla}

\begin{document}

\markboth{S. K. Biswal, S. K. Singh, M. Bhuyan and S. K. Patra}
{Effect of self interacting isoscalar-vector meson on nuclear system}

\catchline{}{}{}{}{}

\title{The effect of self interacting isoscalar-vector meson
on finite nuclei and infinite nuclear matter.}

\author{S. K. Biswal, S. K. Singh, M. Bhuyan and S. K. Patra}
\address{Institute of Physics, Sachivalaya Marg, Bhubaneswar-751 005, India.}

\maketitle

\pub{Received (Day Month Year)}{Revised (Day Month Year)}

\begin{abstract}
A detailed study is made for the nucleon-nucleon interaction based
on relativistic mean field theory in which the potential is explicitly
expressed in terms of mass and the coupling constant of the meson fields.
A unified treatment for self-coupling of isoscalar-scalar $\sigma-$, 
isoscalar-vector $\omega$-mesons and their coupling constant are given
with a complete analytic form. The present investigation is focused on 
the effect of self-interacting higher order $\sigma$ and $\omega$ field 
on nuclear properties. An attempt is made to explain the collapsing
stage of nucleon by higher order $\omega$-field. Both infinite nuclear
matter and the finite nuclear properties are included in the present
study to observe the behaviour or sensitivity of this self interacting
terms.
\end{abstract}

\ccode{21.10.Dr., 21.60.-n., 23.60.+e., 24.10.Jv.}

\section{Introduction}

The Nucleon-Nucleon (NN) interaction problem was started from 
last half century \cite{bethe53,alcock72}. Probably this is a 
long standing question in history of nuclear physics. In fact, 
describing the nuclear properties in terms of the interactions 
between the nucleon’s pairs is indeed the main goal for nuclear 
physicists. The {\it nucleon-nucleon} NN-interaction in terms 
of mediated mesons is put forwarded by Yukawa \cite{yukawa35}
in 1935. Although the meson theory is not fundamental in the 
view of QCD, still the approach has improved our understanding 
of the nuclear forces as well as highlight some good quantitative 
results \cite{fuji86,fuji86a}. The modern theory of NN potential 
through particle exchanges is made possible by the development 
of quantum field theory \cite{fuji86a}. However, at low-energy, 
one can assume that the interactions is instantaneous and 
therefore the concept of interaction potential becomes useful. 
The derivation of a potential through particle exchange, is 
important to understand the nuclear force as well as structural 
properties.

Nowadays, there are large development in the nuclear theory by
introducing quark and gluon in connection with the NN-potential
\cite{gross11,mech11}. These  models give the fundamental
understanding of NN interaction at present. Here, we are not
addressing all these rich and long standing subject about
NN-potential, but some basic facts and important issues of the
NN-interactions arising from relativistic mean field (RMF) 
Lagrangian \cite{bhu12,bhu13,love70,love72}. The behaviour of 
this potential gives an idea about the breaking of nucleon and 
the formation of $quark-gluon-plamsa$ medium at very high energy 
i.e. the collapsing state of nucleons.

This paper is organized as follows. In Section II, we briefly discuss
the theoretical formalism of NN-interaction from relativistic mean
field theory. The general forms of the NN potentials are expressed
in the coordinate space ({\it r-space}) in terms of mass and coupling
constant of the force parameters. In Section III, we review the effect of
modified term in the Lagrangian and their effect on the finite nucleus
and infinite nuclear matter observables. In Section IV, we address few
comments about the current form of the NN-interaction to the collapsing
stage of the nucleons.

\section{The theoretical frameworks}

The nuclear interaction in relativistic mean field is possible via
various mesons interaction with nucleons. The linear relativistic
mean field (RMF) Lagrangian density for a nucleon-meson many-body system
\cite{miller72,walecka74,horo81,ser86} is given as:
\begin{eqnarray}
{\cal L}&=&\overline{\psi_{i}}\{i\gamma^{\mu}
\partial_{\mu}-M\}\psi_{i}
+{\frac12}\partial^{\mu}\sigma\partial_{\mu}\sigma
-{\frac12}m_{\sigma}^{2}\sigma^{2} 
-g_{s}\overline{\psi_{i}}\psi_{i}\sigma \nonumber\\
&&-{\frac14}\Omega^{\mu\nu}
\Omega_{\mu\nu}+{\frac12}m_{w}^{2}V^{\mu}V_{\mu} 
-g_{w}\overline\psi_{i} \gamma^{\mu}\psi_{i} V_{\mu}
-{\frac14}\vec{B}^{\mu\nu}.\vec{B}_{\mu\nu} \nonumber\\
&&+{\frac12}m_{\rho}^{2}{\vec
R^{\mu}} .{\vec{R}_{\mu}} -g_{\rho}\overline\psi_{i}
\gamma^{\mu}\vec{\tau}\psi_{i}.\vec
{R^{\mu}}-{\frac14}F^{\mu\nu}F_{\mu\nu}-e\overline\psi_{i}
\gamma^{\mu}\frac{\left(1-\tau_{3i}\right)}{2}\psi_{i}A_{\mu}, 
\end{eqnarray}
where, the field for $\sigma$ meson is denoted by $\sigma$, for $\omega$
by $V_{\mu}$, and for the iso-vector $\rho$ mesons by $\vec{R}_{\mu}$,
respectively. The $\psi_{i}$, $\tau$ and $\tau_3$ are the Dirac spinors
for nucleons, the iso-spin and the $3^{rd}$ component of the iso-spin,
respectively. Here $g_{\sigma}$, $g_{\omega}$, $g_{\rho}$ and $g_{\delta}$
are the coupling constants for $\sigma$, $\omega$, and $\rho$ mesons and
their masses are denoted by $m_{\sigma}$, $m_{\omega}$ and $m_{\rho}$,
respectively. The field tensors for $V^{\mu}$ and $\vec{R}_{\mu}$ are
given by $\Omega^{\mu\nu}$ and $\vec{B}_{\mu\nu}$, respectively. If, we
neglect the $\rho-$ meson, it corresponds to the Walecka model in its original
form \cite{walecka74,horo81}. From the above relativistic Lagrangian, we
obtain the field equations for the nucleons and mesons as,
\begin{eqnarray}
\Bigl(-i\alpha.\bigtriangledown+\beta(M+g_{\sigma}\sigma)
+g_{\omega}\omega+g_{\rho}{\tau}_3{\rho}_3 
+g_{\delta}\delta{\tau}\Bigr){\psi}_i={\epsilon}_i{\psi}_i,\\
\nonumber \\
(-\bigtriangledown^{2}+m_{\sigma}^{2})\sigma(r)=-g_{\sigma}{\rho}_s(r),\\
\nonumber\\
(-\bigtriangledown^{2}+m_{\omega}^{2})V(r)=g_{\omega}{\rho}(r),\\
\nonumber \\
(-\bigtriangledown^{2}+m_{\rho}^{2})\rho(r)=g_{\rho}{\rho}_3(r),
\end{eqnarray}
for Dirac nucleons and corresponding mesons in the Lagrangian. In the
limit of one-meson exchange and mean-field (the fields are replaced by
their number), for a heavy and static baryonic medium, the solution of
single nucleon-nucleon potential for scalar ($\sigma$) and vector
($\omega$, $\rho$) fields are given by \cite{ser86,brock78},
\begin{eqnarray}
V_{\sigma}(r)&=&-\frac{g_{\sigma}^{2}}{4{\pi}}\frac{e^{-m_{\sigma}r}}{r}, 
\end{eqnarray}
and
\begin{eqnarray}
V_{\omega}(r)&=&+\frac{g_{\omega}^{2}}{4{\pi}}\frac{e^{-m_{\omega}r}}{r},\quad
V_{\rho}(r)=+\frac{g_{\rho}^{2}}{4{\pi}}\frac{e^{-m_{\rho}r}}{r}.
\end{eqnarray}
The total effective nucleon-nucleon potential is obtained from the scalar
and vector parts of the meson fields. This can be expressed as \cite{bhu12},
\begin{eqnarray}
v_{eff}(r)&=&V_{\omega} +V_{\rho} +V_{\sigma} 
=\frac{g_{\omega}^{2}}{4{\pi}}\frac{e^{-m_{\omega}r}}{r}
+\frac{g_{\rho}^{2}}{4{\pi}}\frac{e^{-m_{\rho}r}}{r}
-\frac{g_{\sigma}^{2}}{4{\pi}}\frac{e^{-m_{\sigma}r}}{r}.
\label{eq:10}
\end{eqnarray}

\subsection{Non-linear case}
The Lagrangian density in the above Eqn. (1) contains only linear
coupling terms, which is able to give a qualitative description of the
nuclei \cite{ser86,brock78}. The essential nuclear matter properties like
incompressibility and the surface properties of the finite nuclei cannot
be reproduced quantitatively within this linear model. Again the
interaction between a pair of nucleons when they are embedded in a heavy
nucleus is less than the force in empty space. This suppression of
the two-body interactions within a nucleus in favour of the interaction
of each nucleon with the average nucleon density, means that the
non-linearity acts as a smoothing mechanism and hence leads in the
direction of the one-body potential and shell structure
\cite{boguta77,bodmer91,sgmuca92,suga94}. The replacement of mass
term $\frac{1}{2} m_{\sigma}^2\sigma^2$
of $\sigma$ field by $U (\sigma)$ and
$\frac{1}{2} m_{\omega}^2 V^{\mu} V_{\mu}$ of $\omega$ field by
$U (\omega)$. This can be expressed as
\begin{eqnarray}
U (\sigma) = \frac{1}{2}m_{\sigma}^2 \sigma^2+ \frac{1}{3}g_{2}\sigma^{3} 
+\frac{1}{4} g_{3}\sigma^{4}, \\ 
U (\omega) = \frac{1}{2}m_{\omega}^2 V_{\mu}V^{\mu} +\frac{1}{4}c_3 (V_{\mu}V^{\mu})^{2}.
\end{eqnarray}
The terms on the right side of Eqns (9-10), except the first term
other are from the non-linear self coupling amongst the $\sigma$ and
$\omega$ mesons, respectively \cite{boguta77,bodmer91}. Here, the
non-linear parameter $g_2$ and $g_3$ due to $\sigma-$ fields are adjusted
to the surface properties of finite nuclei \cite{schif51,schiff051}.
The most successful fits yield, the $+ve$ and $-ve$ signs for $g_2$ and
$g_3$, respectively. The negative value of $g_3$ is a serious problem in
quantum field theory. As, we are dealing within the mean field level and
with normal nuclear matter density, the corresponding $\sigma$ field is
very small and the $-ve$ value of $g_3$ is still allowed
\cite{schif51,brock90}. With the addition of the non-linear terms of the
Eqns (9-10) to the Lagrangian, the field equation for $\sigma$ and
$\omega$- fields (in Eqn. (6-7)) are modified as:
\begin{eqnarray}
(-\bigtriangledown^{2}+m_{\sigma}^{2})\sigma(r)=-g_{\sigma}{\rho}_s(r)
-g_2\sigma^2 (r) - g_3 \sigma^3 (r), \nonumber \\
\nonumber \\
(-\bigtriangledown^{2}+m_{\omega}^{2})V(r)=g_{\omega}{\rho}(r)
-c_3 W^3 (r).
\end{eqnarray}
Here, $W(r)=g_{\omega}V_0 (r)$ and $c_3$ is the non-linear coupling
constant for self-interacting $\omega$-mesons. Because of the great
difficulty in solving the above
nonlinear differential equations, it is essential to have a variation
principle available for the estimation of the energies associated
with various source distributions. In the static case, the negative
sign of the third term in the Lagrangian is computed with the correct
source function and an arbitrary trial wave function. The limit on the
energy has a stationary value equal to the correct energy when the
trial function is in the infinitesimal neighborhood of the correct
wave function. Now, the solution for the modified $\sigma$ and $\omega$
fields are given as \cite{schif51}
\begin{eqnarray}
V_{\sigma}=-\frac{g_{\sigma}^{2}}{4{\pi}}\frac{e^{-m_{\sigma}r}}{r}
+\frac{g_{2}^{2}}{4{\pi}}\frac{e^{-2m_{\sigma}r}}{r^2}
+\frac{g_{3}^{2}}{4{\pi}}\frac{e^{-3m_{\sigma}r}}{r^3}, \nonumber \\
\nonumber \\
V_{\omega}=-\frac{g_{\omega}^{2}}{4{\pi}}\frac{e^{-m_{\omega}r}}{r}
+\frac{c_{3}^{2}}{4{\pi}}\frac{e^{-3m_{\omega}r}}{r^2}.
\label{eq:10}
\end{eqnarray}
The new NN-interaction analogous to $M3Y$ form and is able to improve
the incompressibility and deformation of the finite nuclei results
\cite{brock90}. In addition to this, the non-linear self coupling
of the $\sigma$ and $\omega$-mesons help to generate the repulsive
and attractive part of the NN potential at $long$ as well as at $short$
distance respectively to satisfy the saturation properties (Coester-band
problem) \cite{schiff051}. Thus also, generate the most discussed $3-body$
interaction. The modified effective nucleon-nucleon interaction is
defined as \cite{bhu12}:
\begin{eqnarray}
v_{eff}(r)&=&V_{\omega}+V_{\rho}+V_{\sigma}\nonumber \\
&& =\frac{g_{\omega}^{2}}{4{\pi}}\frac{e^{-m_{\omega}r}}{r}
+\frac{g_{\rho}^{2}}{4{\pi}}\frac{e^{-m_{\rho}r}}{r}
-\frac{g_{\sigma}^{2}}{4{\pi}}\frac{e^{-m_{\sigma}r}}{r} \nonumber \\
&& +\frac{g_{2}^{2}}{4{\pi}}\frac{e^{-2m_{\sigma}r}}{r^2}
+\frac{g_{3}^{2}}{4{\pi}}\frac{e^{-3m_{\sigma}r}}{r^3}
+\frac{c_{3}^{2}}{4{\pi}}\frac{e^{-3m_{\omega}r}}{r^3}
\label{eq:10}
\end{eqnarray}
\begin{table}
\caption{The values of $m_{\sigma}$, $m_{\omega}$, $m_{\rho}$ 
(in MeV) and $g_{\sigma}$, $g_{\omega}$, $g_{\rho}$ for RMF (NL3) 
force, along with the self-interacting $\omega-$ field with coupling 
constant $c_3$ [26,27].}
\renewcommand{\tabcolsep}{0.15cm}
\renewcommand{\arraystretch}{1.5}
\begin{tabular}{|cccccccccc|ccccccc}
\hline
Set&$m_{\sigma}$& $m_{\omega}$ & $m_{\rho}$& $g_{\sigma}$ &$g_{\omega}$&$g_{\rho}$&
$g_2$ & $g_3$ &$c_3$\\
\hline
NL3 & 508.194 & 782.5 & 763.0 & 08.31 & 13.18 & 6.37 & -10.4307 & -28.8851 & 0.0 $\pm$ 0.6 \\
\hline
\end{tabular}
\label{Table 1}
\end{table}

\section{Results and Discussion}
The above expression in Eqn. (13) shows that the effective nucleon-nucleon
potential is presented eloquently in terms of the well known inbuilt RMF
theory parameters of $\sigma$, $\omega$ and $\rho$ meson fields. Here,
we have used RMF (NL3) force parameter along with varying $c_3$ for
$\omega$-self interactions to determine the nuclear properties. The
values of these ardent parameter for NL3-force are listed in Table 1.
Although, the $\omega^4$ term is already there in the FSU-Gold parameter
\cite{pika01,pika05}, here we are interested to see the effect
of non-linear self coupling of $\omega$ meson. Thus, we have added
the self-interaction of $\omega$ with coupling constant $c_3$ on the
top of NL3 sets and observing the possible effects.

\begin{figure}
\vspace{1.0cm}
\includegraphics[width=0.8\columnwidth]{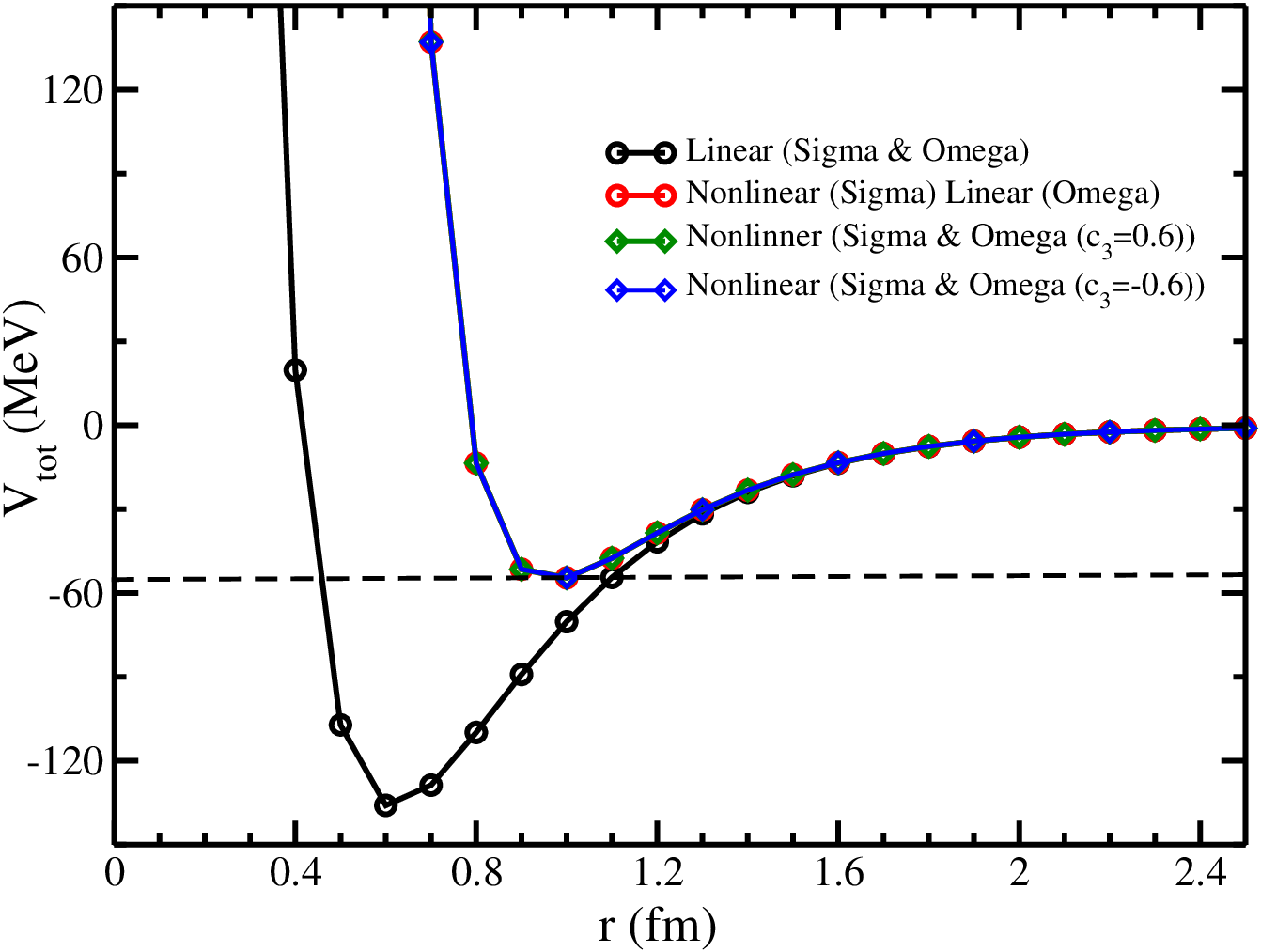}
\caption{The effective NN interaction potentials as a function of
distance $r$ from Eqn. (7-10) for NL3 parameter set.}
\label{Fig. 1}
\end{figure}

First of all, we have calculated the NN-potential for linear and
non-linear cases using Eqns (8) and (13), respectively. The obtained
results for each cases are shown in Fig. 1. From the figure, it is
clear that without taking the non-linear coupling for RMF (NL3), one
cannot reproduce a better NN-potential. In other word, the depth of
the potential for linear and non-linear are $\sim$ 150 MeV and 50 MeV,
respectively. Thus, the magnitude of the depth for linear case is not
reasonable to fit the NN-data. Again, considering the values of $c_3$,
there is no significant change in the total {\it nucleon-nucleon}
potential. For example, the NN-potential does not change at all for
$c_3$ $\simeq$ $\pm$ 0.6, which can be seen from Fig. 1.

\begin{figure}
\vspace{1.0cm}
\includegraphics[width=0.8\columnwidth]{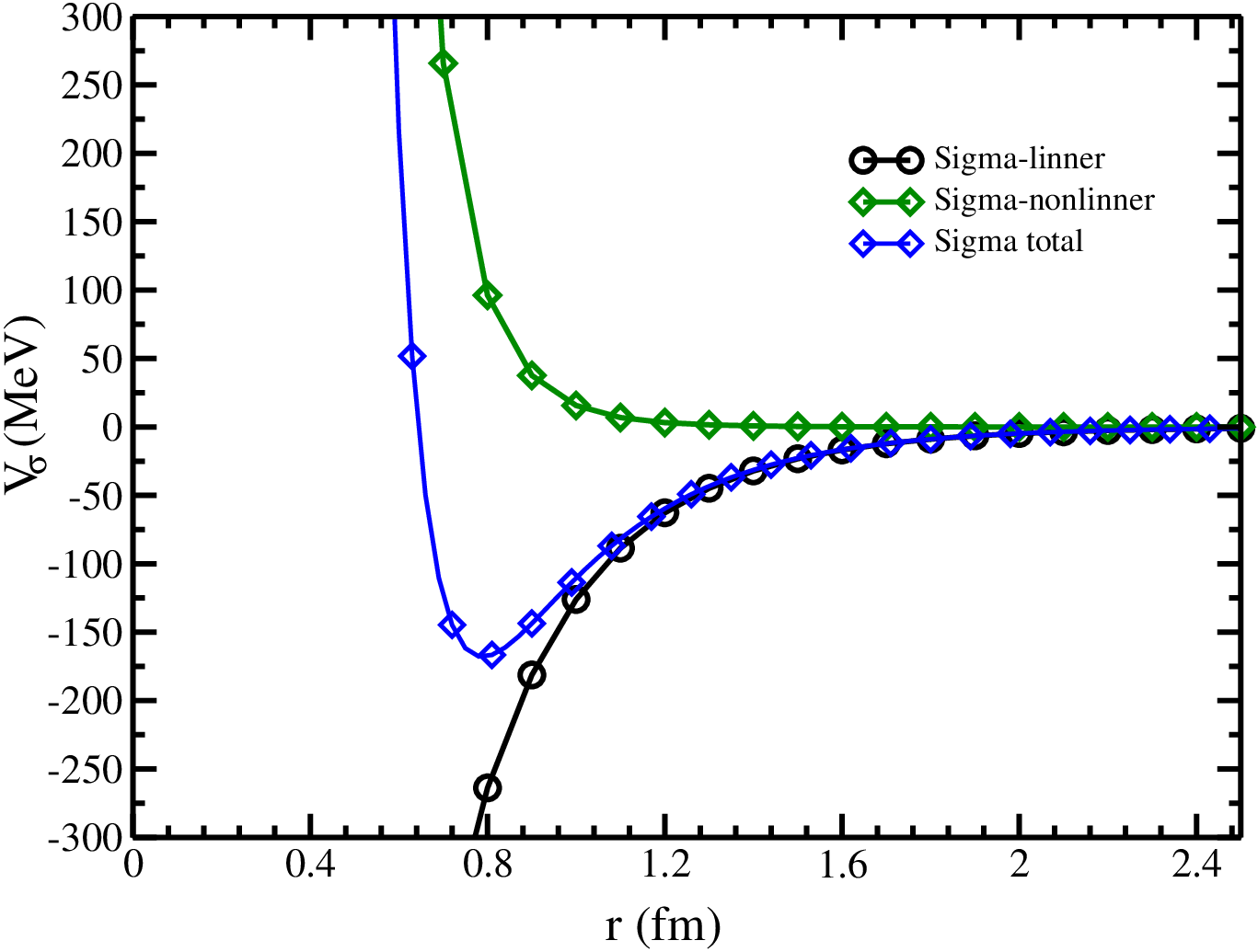}
\caption{The contribution of $\sigma$-potential from linear, non-linear
and total as a function of distance $r$ for for NL3 parameter set.}
\label{Fig. 1}
\end{figure}

\begin{figure}
\vspace{1.0cm}
\includegraphics[width=0.8\columnwidth]{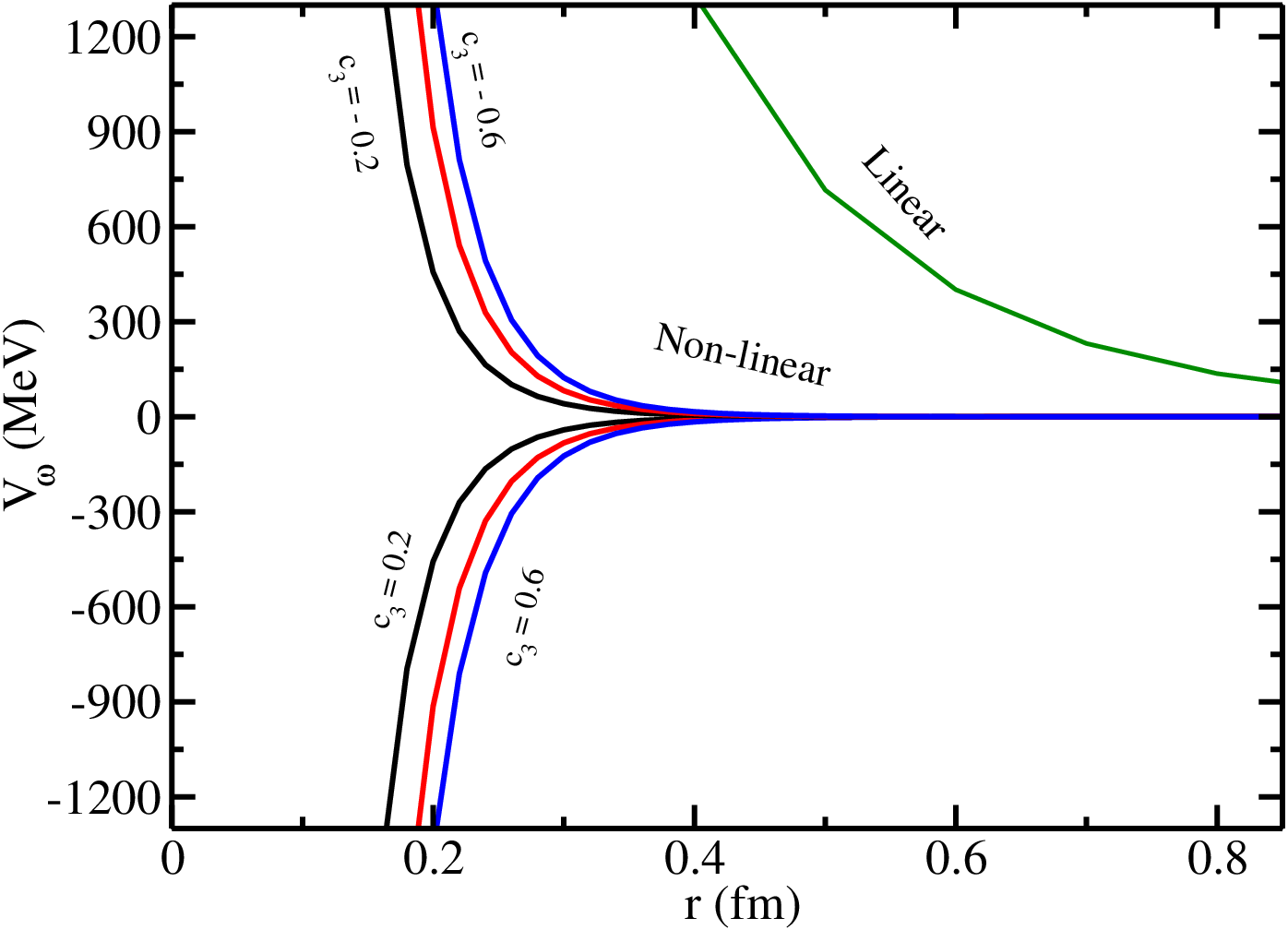}
\caption{The contribution of $\omega$-potential from linear, non-linear
and total as a function of distance $r$ for for NL3 parameter set.
}
\label{Fig. 3}
\end{figure}

Further, we have calculated the individual contribution of meson
fields to the NN-potential in particular case of $\sigma$ and
$\omega$-mesons. In case of $\sigma$-field, we have calculated
the linear and non-linear contribution separately, and combined
to get the total $\sigma$-potential, which is shown in Fig. 2.
From the figure, one can find the non-linear self-interacting
terms in the $\sigma$-field play an important role in the
repulsive core of the total NN-potential \cite{schif51}. The
linear and non-linear contribution of the $\omega$-field at
various $c_3$ are shown in Fig. 3. The important feature in this
figure is that the linear term give an infinity large repulsive
barrier at $\sim 0.4$ fm, at which range, the influence of the non-linear
term of the $\omega-$meson is till zero. However, this non-linear
terms is extremely active at very short distance ($\sim 0.2$ fm), which
can be seen from the figure. Further, for both $\pm$-values
of $c_3$, the contribution having same magnitude but in opposite
directions. The strongly attractive potential of the non-linear
$\omega-$ meson part towards the central region make some evidence
to explain the collapsing stage of the nucleons during the formation
of $quark-gluon-plasma$ QGP at high energy heavy-ion-collision. That
means, mostly the (i) $\sigma-$
meson is responsible for the attractive part of the nuclear force
(nuclear binding energy) (ii) the non-linear terms are responsible
for the repulsive part of the nuclear force at long distance, which
simulate the 3-body interaction of the nuclear force \cite{schif51}.
(help to explain the Coester band problem) (iii) similarly,
the $\omega-$ meson is restraint for the repulsive part of the
nuclear force ({\it known as hard core}) and (iv) the non-linear
self-coupling of the $\omega-$ meson ($\frac{1}{4}c_3 V_{\mu}V^{\mu}$
is responsible for the attractive part in the very shortest
($\sim 0.2 fm$) region of the NN-potential. This short range strong
attractive component of the $\omega-$ meson is important for the
collapsing stage of the nucleons, when they come close to each other
and overcome the barrier during the heavy-ion-collision (HIC)
experiment and makes the QGP state of matter. The shifting of the
barrier towards the central region make some evidence to
explaining the collapsing state of the nucleon. It is worthy to
mention that the values of these constants are different for
different forces of RMF theory. Hence, the NN-potential somewhat
change a little bit in magnitude by taking different forces,
but the nature of the potential remains unchanged.

\begin{figure}
\vspace{1.0cm}
\includegraphics[width=0.8\columnwidth]{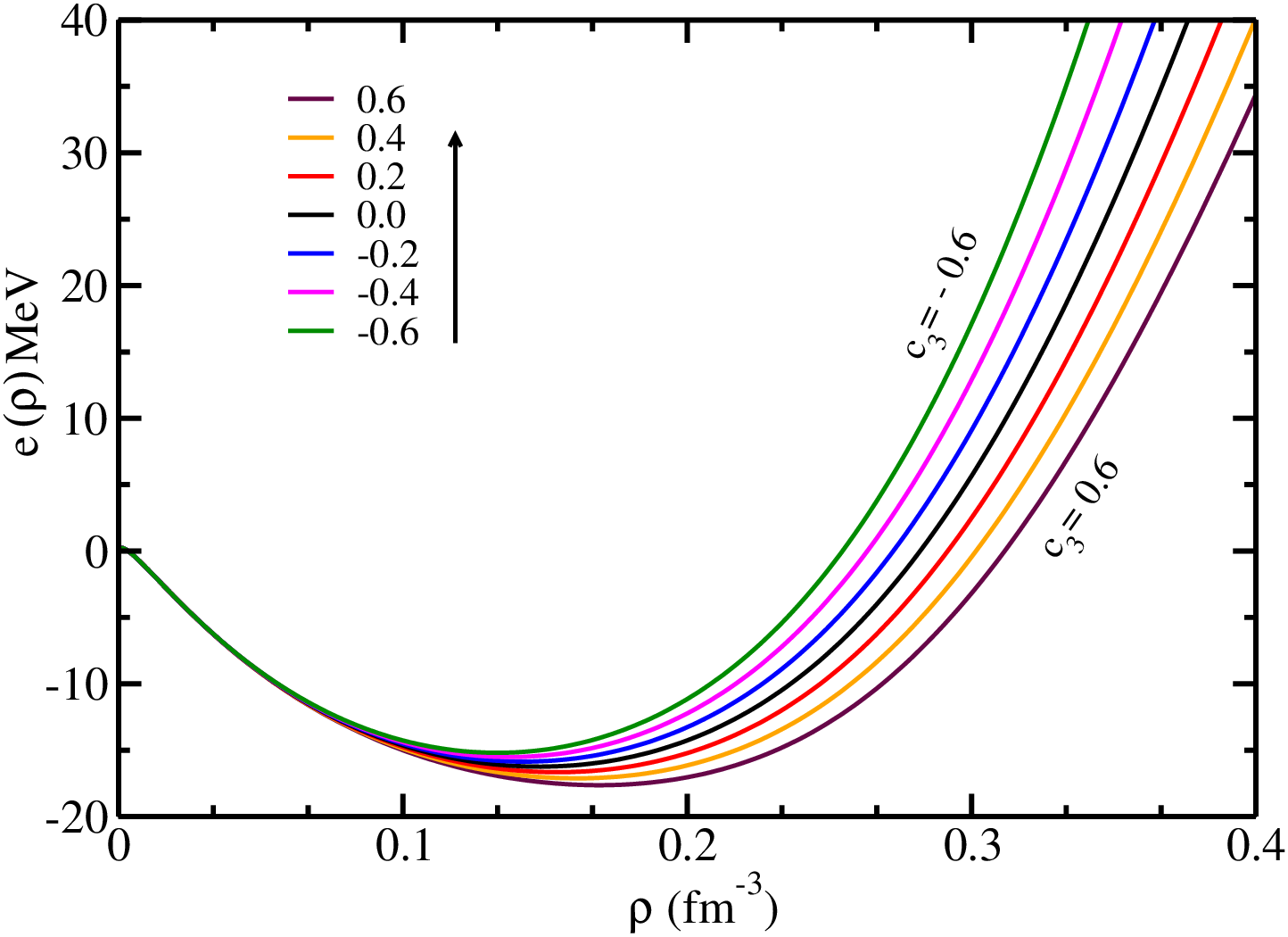}
\caption{The energy per particle of symmetric nuclear matter as a function
of baryon density for various values of $c_3$.}
\label{Fig. 6}
\end{figure}

\subsection{Energy density and Pressure density}

In the present work, we study the effect of the additional term
on top of the $NL3$ force parameter to the Lagrangian, which comes
from the self-interaction of the vector fields with $c_3$ as done
in the Refs. \cite{pika01,pika05}. The inclusion of this term is
not new, it is already taken into account for different forces
of RMF and effective field theory motivated relativistic mean
field theory (E-RMF). Here, our aims to see the effect of $c_3$ to
the nuclear system and the contribution to the attractive part of
the hard core of NN-potential. We have solved the mean field equations
self-consistently and estimated the energy and pressure density as
a function of baryon density. The NL3 parameter set \cite{furn87,furn89}
along with the additional $c_3$ is used in the calculations. The obtained
results for different values of $c_3$ are shown in Figs. 4 and 5,
respectively. From the figure, it is clearly identify that the $-ve$
value of $c_3$ gives the {\it stiff} equation of state (EOS), meanwhile
the $+ve$ value shows the {\it soft} EOS. It is to be noted that mass
and radius of the neutron star depends on the softness and stiffness
of EOS. Here, in our investigation, we observed that the softening
of the EOS depends on the non-linear coupling of the $\omega-$ meson
\cite{pika05}. The recent measurement of Demorest {\it et. al.}
\cite{demo10} put a new direction that the NL3 force needs slightly
softer EOS. However, when we deals with G2 (E-RMF) model, the results
of the Ref. \cite{singh13} demands a slightly stiffer EOS. This implies
that, the value of $c_3$ should be fixed according to solve the above
discussed problem.

\begin{figure}
\vspace{1.0cm}
\includegraphics[width=0.8\columnwidth]{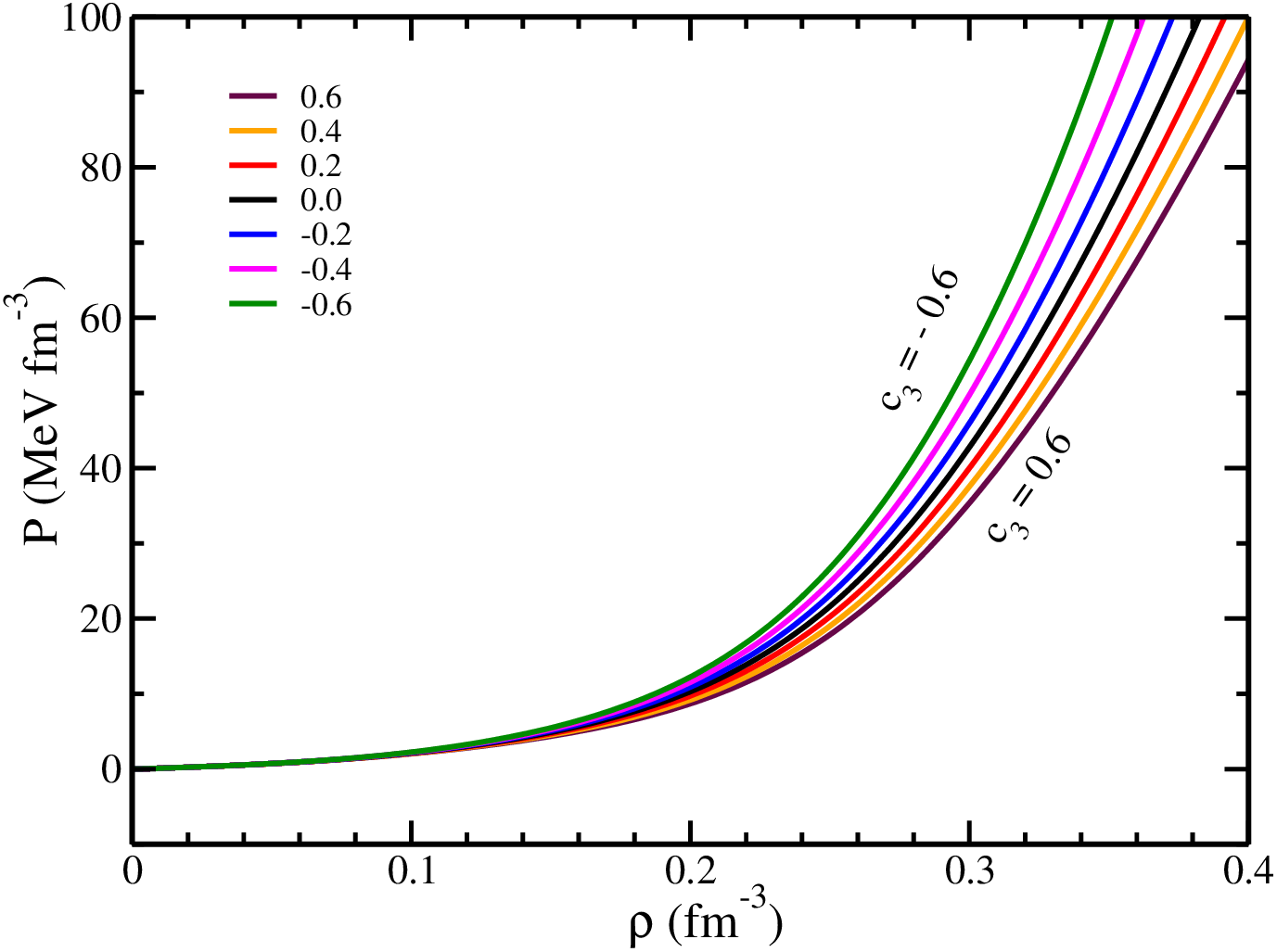}
\caption{The pressure density of symmetric nuclear matter as a function
of baryon density for various values of $c_3$.}
\vspace{-0.2cm}
\label{Fig. 8}
\end{figure}

\begin{figure}
\vspace{1.0cm}
\includegraphics[width=0.8\columnwidth]{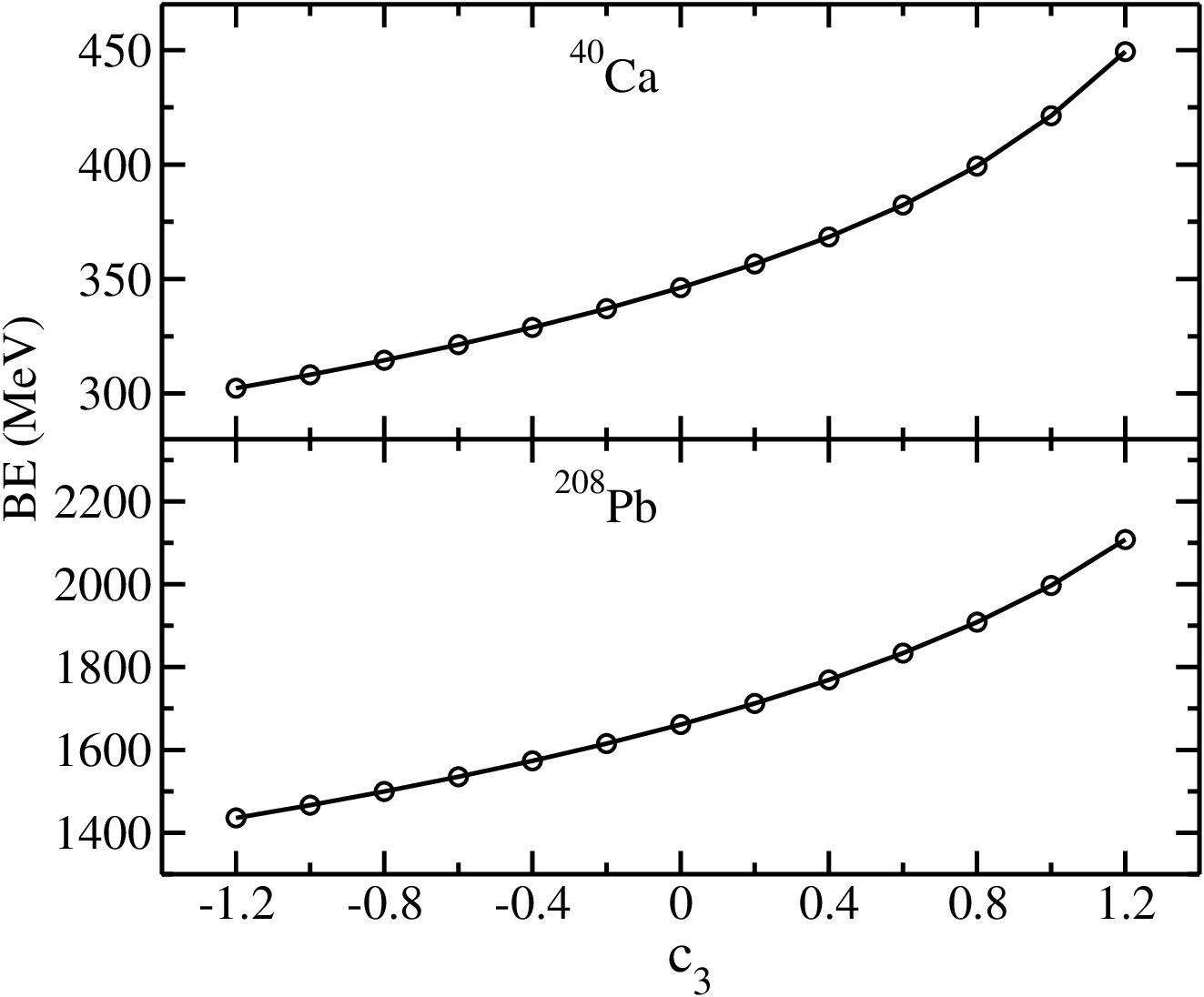}
\caption{The binding energy of $^{40}$Ca and $^{208}$Pb in their ground
state for different values of $c_3$.}
\label{Fig. 5}
\end{figure}

\begin{figure}
\vspace{1.0cm}
\includegraphics[width=0.8\columnwidth]{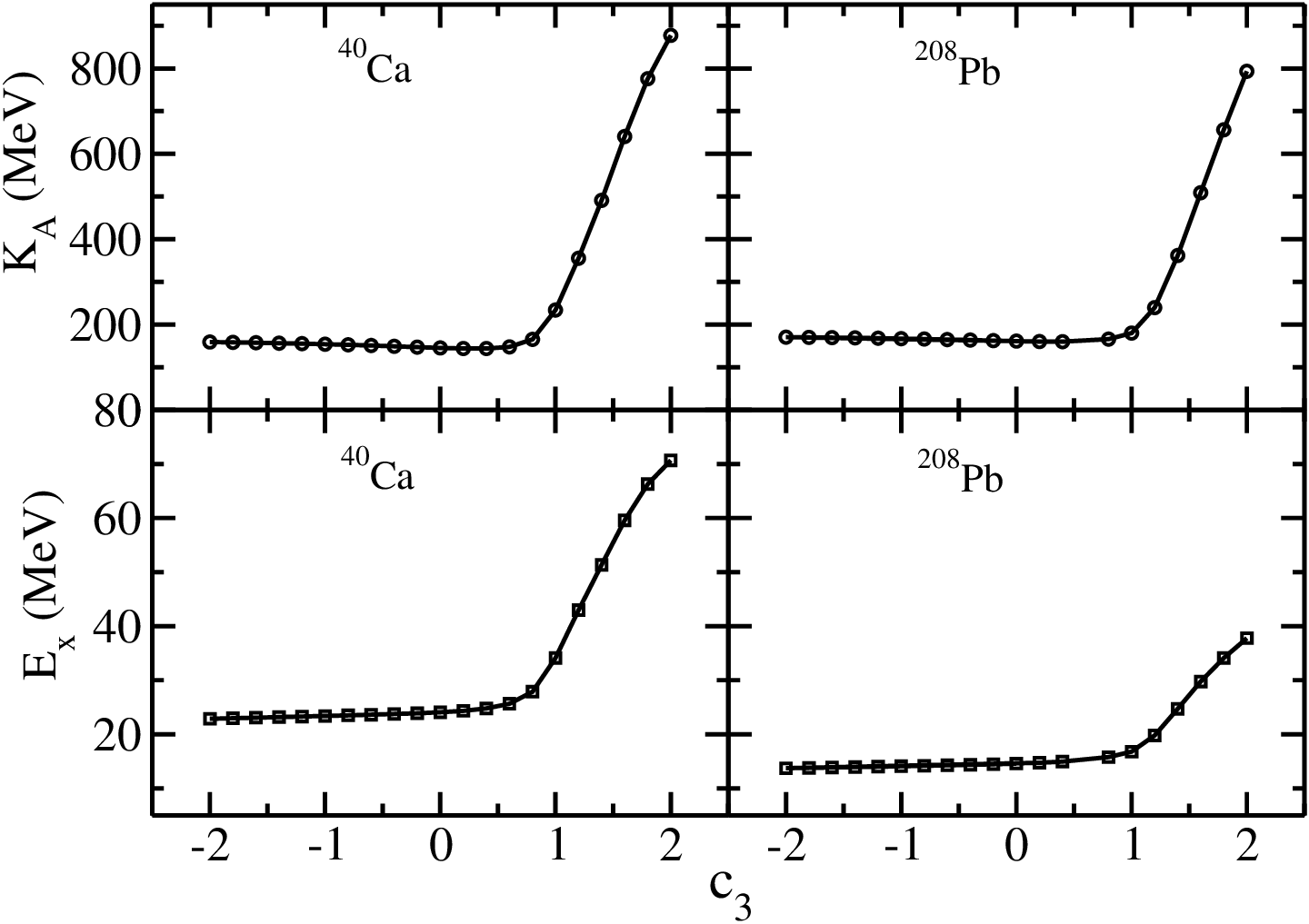}
\caption{(a) The excitation energy as a function of $c_3$ for $^{40}$Ca
and $^{208}$. (b) The incompressibility of $^{40}$Ca and $^{208}$ as
a function of $c_3$.}
\end{figure}

\subsection{Binding energy, Excitation energy and Compressibility}
To see the sensitivity of $c_3$ on the finite nuclei, we calculated
the binding energy (BE), giant monopole excitation energy ($E_x$)
for $^{40}$Ca and $^{208}$Pb nuclei as representative cases as
a function of $c_3$. The obtained results are shown in upper and
lower panel of the Fig. 6, respectively. From the figure, one can
observed a systematic variation of binding energy by employing
the isoscalar-vector self-coupling parameter $c_3$. For example,
the binding energy is monotonically changes for all $\pm ve$
values of $c_3$.

Further, we analyzed the variation of compressibility
modulus of $^{40}{Ca}$ and $^{208}{Pb}$ in the upper panel of
the Fig. 7 and excitation energy of these nuclei in the lower
panel.The value of the coupling constant($c_3$) varies from
$-2$ to $+2$, where the result have reasonable values. We included
both positive and negative value of $c_3$ to know the the
discrepancy between the sign of $c_3$. Excitation energy and
compressibility modulus in finite nuclei are calculated by using
scaling method and the extended Thomas-Fermi approaches to relativistic
mean field theory (RETF) \cite{subrat}. From the calculated results,
we find that the compressibility modulus as well as the monopole
excitation energy of finite nuclei do not change with the increase
of $c_3$ to some optimum value. It is interesting to notice that
although we get a stiff equation of state with negative value of
$c_3$ for infinite nuclear matter system, this behaviour does not
result in finite nuclei, i.e. the $K_A$ and $E_x$ do not change for
negative $c_3$. May be the density with which we deal in the finite 
nucleus is responsible for this discrepancy. However, $K_A$ and $E_x$ 
increase substantial after certain value of $c_3$, i.e. the finite 
nucleus becomes too much softer at about $c_3 \sim$ 1.0. As a result, 
the incompressibility becomes large.

\section{Summary and Conclusions}

In this paper, we tested the effect of the non-linear self-coupling
of the $\omega-$ vector meson. At extremely short distance, it gives
a strongly attraction for $-ve$ value of $c_3$, which is mostly
responsible for the asymptotic properties of the quarks. This short
range distance is about $0.2 fm$, below which the $vector-$meson itself
shows a strong attraction due to its self-interaction. This short-range
strong attraction makes the nucleon collapse and form $quark-gluon-plasma$
when a highly energetic projectile nucleon approaches another target nucleon
of a distance of $\sim 0.2$ fm.

\end{document}